\documentclass[11pt]{paper}

\usepackage{amssymb}
\usepackage{amscd}
\usepackage{amsfonts}
\usepackage{amssymb}
\usepackage{cite}
\usepackage{amsmath}
\usepackage{epsfig}
\usepackage{graphicx, subfigure}
\usepackage{color, graphics}
\usepackage{pdfsync}
\usepackage{hyperref}
\usepackage[all,cmtip]{xy}
\usepackage{setspace}
\usepackage[all,cmtip]{xy}
\usepackage{tikz}
\usepackage{tikz-cd}
\usepackage{pst-all}
\usepackage{pstricks-add}
 \newtheorem{thm}{\textbf{Theorem}}[section]

 \newtheorem{cor}{\textbf{Corollary}}[section]
 \newtheorem{rem}{\textbf{Remark}}[section]
 \numberwithin{equation}{section}

\renewcommand{\le}{\leqslant}
\renewcommand{\ge}{\geqslant}

\allowdisplaybreaks

\newcommand{\R}{\mathbb{ R}}

\DeclareMathOperator{\pr}{pr}
\DeclareMathOperator{\Span}{span}
\DeclareMathOperator{\diag}{diag}

\DeclareMathOperator{\tr}{tr}

\def\diag{\mathrm{diag}}

\def\px1{p_{x_1}}
\def\px2{p_{x_2}}
\def\pu1{p_{u_1}}

\begin{document}

\title{The Zhukovskiy--Volterra gyroscope: geometry and reduction}

%\keywords{Liouville integrability; Lax representation, noncommutative integrability; heavy rigid body; gyroscope}

\author{Vladimir Dragovi\'c, Borislav Gaji\'c, and Bo\v zidar Jovanovi\'c}

\maketitle

\noindent{\small Department of Mathematical Sciences, The University of Texas at Dallas,  USA \& Mathematical Institute SANU, Belgrade, Serbia
\footnote{{\sc email:} Vladimir.Dragovic@utdallas.edu}}

\noindent{\small Mathematical Institute SANU, Belgrade,
Serbia\footnote{{\sc email:} gajab@mi.sanu.ac.rs}$^,$\footnote{{\sc email:} bozaj@mi.sanu.ac.rs}}

\begin{abstract}
We describe a symplectic reduction of the Zhukovskiy integrable case of motion of a rigid body about a fixed point with a gyroscope in $\R^n$ to an integrable magnetic flow on a sphere $S^{n-1}$.  We also present Zhukovskiy's geometric interpretation of motion of the Euler
top with a gyroscope, aka the Zhukovskiy--Volterra gyroscope,  and perform a symplectic reduction of this system to an integrable magnetic flow on a $2$-dimensional sphere.
\end{abstract}

\maketitle

\section{Introduction}

The Euler case of motion of a rigid body with a gyroscope about the fixed point is considered by Zhukovskiy in 1885 (see \cite{Zh}).
 The equations of motion of a rigid body with a gyroscope of a constant angular momentum $\mathbf L=(L_1,L_2,L_3)$ in the body frame are given by
\begin{equation}\label{MEP}
\dot{\mathbf M}=\big(\mathbf M+\mathbf L\big) \times \Omega,  \qquad \mathbf M=I\Omega,
\end{equation}
where $\Omega$ is the angular velocity of the body in the moving frame,
and $I$ be the operator of inertia of the  system body+gyroscope (see Section \ref{sec2}).

 Voltera integrated the equations of motion \eqref{MEP} in 1899 in \cite{Vol} (see also \cite{Bas1, Bas3, W}).  The Euler case with a gyroscope is usually called \emph{Zhukovskiy--Volterra gyroscope}.
Further analysis of this system was performed, for example in \cite{Bas2, Bas3, BM2019, Khar, GK}.

Prior to  Volterra's solution from \cite{Vol},
Bobilev  solved the equations of motion of a nonholonomic problem of a
 balanced ball, where balanced means that the geometric center coincides with the mass center, with a gyroscope rolling without slipping over a plane in 1892 in \cite{Bob}.
Bobilev's  assumption was that the central ellipsoid of inertia of the ball is rotationally symmetric
and the gyroscope axis $\ell$ coincides with the axis of symmetry of the ball.
 Zhukovskiy (1893) noticed  that the reduction to the elliptic quadratures would be simpler with an additional assumption, \emph{called the Zhukovskiy condition},
that $I$ is proportional to the identity  matrix (see \cite{Zhuk1893}).
Demchenko used the Zhukovskiy condition to integrate the equations of motion of the problem of rolling of a gyroscopic ball over a sphere in his PhD thesis in 1923 \cite{Dem1924} (see also \cite{DGJ}).

We considered an $n$--dimensional generalization of the Demchenko case in \cite{DGJ2023}. This is a nonholonomic problem of rolling of a ball with a gyroscope without slipping and twisting over a hyperplane and over a sphere in $\R^n$, with the inertia operator of the system ball+gyroscope being proportional to the identity operator on $so(n)$ -
the Zhukovskiy condition in $\R^n$ from \cite{DGJ2023}.
This system is an example of a $G$-Chaplygin system with gyroscopic forces and allows for a reduction to the magnetic geodesic flow on the sphere
\begin{equation}\label{sfera}
S^{n-1}=\{\gamma\in\R^n\,\vert\, \langle \gamma,\gamma\rangle=1\}
\end{equation}
placed in the homogeneous magnetic field $s\kappa$ in $\R^n$  (see \cite{DGJ2025}):
\begin{equation}\label{stari}
\ddot \gamma=
{s}\kappa \dot \gamma +\lambda\gamma, \qquad \lambda={s}\langle \dot\gamma,\kappa\gamma \rangle-\langle \dot\gamma,\dot\gamma\rangle.
\end{equation}
Here $\kappa$ is a fixed skew-symmetric matrix, $s$ is a real papameter depending on the strength of the field, and
the multiplier  $\lambda$ is determined from the  constraint $\langle \gamma(t),\gamma(t)\rangle=1$.

In this paper, following \cite{DGJ2026}, we consider  the motion of a rigid body with a gyroscope  about a fixed point in $\R^n$ and, in particular, the integrable case satisfying the Zhukovskiy condition. In Section \ref{sec3} we perform a Hamiltonian symplectic reduction of the Zhukovskiy integrable case  to another, slightly different from \eqref{stari} magnetic flow on the sphere \eqref{sfera} (see Theorem \ref{prva}):
\begin{equation}\label{novi}
\ddot \gamma=s\kappa\dot\gamma-\frac{1}{4}s^2\kappa^2\gamma+\mu\gamma, \qquad \mu=\frac{1}{4}s^2\langle \kappa^2\gamma,\gamma\rangle+
{s}\langle \dot\gamma,\kappa\gamma \rangle-\langle \dot\gamma,\dot\gamma\rangle.
\end{equation}

By using a Lax representation, we prove that the magnetic flow \eqref{novi} is completely integrable (Theorems \ref{druga} and \ref{treca}, Section \ref{sec3}).
For $n=3$, we also describe a reduction of the Zhukovskiy--Volterra gyroscope to a two-dimensional sphere (Theorems \ref{cetvrta} and \ref{peta}, Section \ref{sec4}). For $s=0$, from Theorem \ref{peta}, we reobtain the well known expression for the Poisson sphere metric \cite{BKF}.

Further, in Section \ref{sec2}, we  present a geometric construction, which goes back to Zhukovskiy in \cite{Zh} and describes the Euler case of  motion of a rigid body with a gyroscope, in transparent geometric terms.

\section{The Zhukovskiy geometric interpretation}\label{sec2}

The rigid body with a gyroscope is a system consisting of two rigid bodies. The second one is symmetric and rotates about the fixed axis of symmetry
$\ell$ and it is called \emph{the gyroscope} in the first rigid body.

 Zhukovskiy  assumed that the center of masses of the gyroscope coincides with the fixed point $O$, while
  the axis of symmetry of the gyroscope $\ell$  does not need to coincides with any of the principal axes of inertia of the rigid body.
   Let $C_2$ be  the moment of inertia of the gyroscope with respect to its axis $\ell$, and  $A_2$ be the moment of inertia of the gyroscope with respect to the line that contains $O$ and lies in the plane $\sigma$, which is orthogonal  to the axis of the gyroscope. Using the symmetry of the gyroscope, one can easily see that the moment of inertia for an arbitrary line in the plane $\sigma$ that contains $O$ has the moment of inertia equal to $A_2$.
    No additional external forces act on the gyroscope.

   Whence, the
angular velocity  $\Omega'_3$ of the gyroscope with respect to its axis is constant  (see \cite{Zh}).

We consider three parts of the  total angular momentum of the system body plus gyroscope $\mathbf{K}$: the angular momentum of the body $\mathbf{M}_0$, and two components of the angular momentum of the gyroscope. These two components of the angular momentum of the gyroscope are: (i) the angular momentum $\mathbf{L}=C_2\Omega'_3\mathbf{E}'_3$ with respect to the gyroscope axis  $\ell$, and (ii) $\mathbf{L}'$ that lies in the plane $\sigma=\ell^\perp$. Here $\mathbf E_3'$ is the unit vector of $\ell$.

Zhukovskiy  defined \emph{the transformed body}  by gluing to the first rigid body an infinite thin  rod lying on the gyroscope axis and having the moment of inertia equal to $A_2$ with respect to the line perpendicular to the axis of the gyroscope that belongs to the plane $\sigma$.  In such a way, the angular momentum of the transformed body is
$$\mathbf{M}=\mathbf{M}_0+\mathbf{L}'=I\Omega.$$
Here $I$ is the new inertia operator, that we call \emph{the inertia operator of the system body+gyroscope}. The total angular momentum of the system is
\[
\mathbf K=\mathbf M+\mathbf L=I\Omega+\mathbf L.
\]
The constant angular momentum $\mathbf L$ is called the \emph{angular momentum of the gyroscope}.

\begin{figure}[h]\label{fig1}
{\centering{\includegraphics[width=5cm]{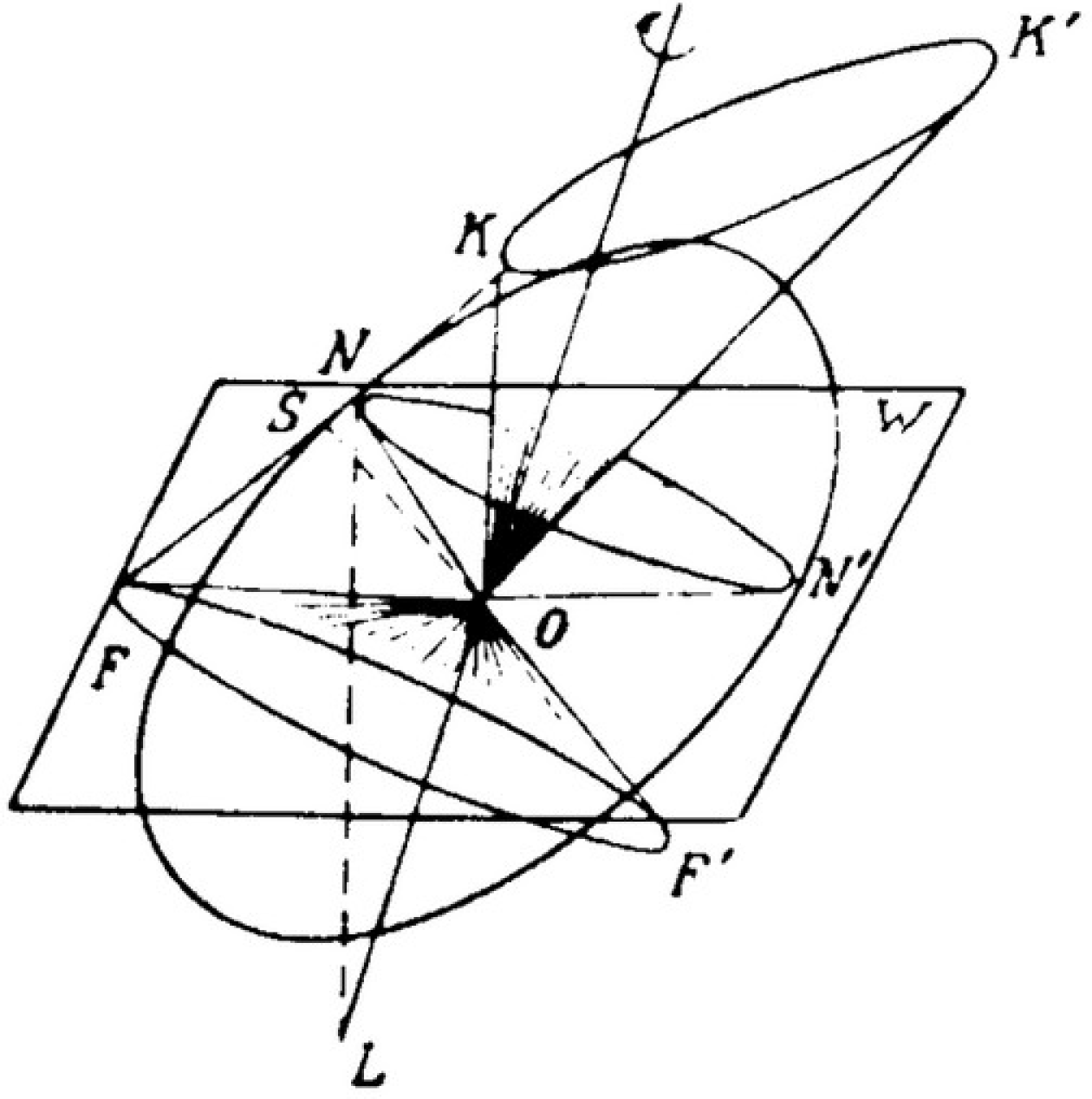}}
\caption{The Zhukovskiy geometric interpretation:  an original figure from \cite{Zh}.}}
\end{figure}

Since the momentum $\mathbf K$ is preserved in the space frame, the equations of motion in the moving frame are \cite{Zh}
\begin{equation}\label{MEP1}
 \dot{\mathbf K}=\dot{\mathbf M}=\big(\mathbf M+\mathbf L\big) \times \Omega=\mathbf K\times\Omega, \qquad \Omega=I^{-1}\mathbf M =I^{-1}\mathbf K-I^{-1}\mathbf L.
\end{equation}

Let $\mathbf E_1$, $\mathbf E_2$, and  $\mathbf E_3$ be the principal  axes of inertia of the transformed body in which $I=\diag(A,B,C)$ and $\mathbf L=(L_1,L_2,L_3)$.
The  kinetic energy $H=h$ and the intensity $\vert\mathbf K\vert=k$ of the total angular momentum of the system body+gyroscope are first integrals of  motion:
\begin{equation}\label{eq:Zhint}
\begin{aligned}
&A\Omega_1^2+B\Omega_2^2+C\Omega_3^2=2h,\\
&(A\Omega_1+L_1)^2+(B\Omega_2+L_2)^2+(C\Omega_3+L_3)^2=k^2.
\end{aligned}
\end{equation}

The starting point of the Zhukovskiy geometric interpretation is the ellipsoid,  that Zhukovskiy referred  to  as
\emph{the MacCullagh ellipsoid of inertia} \cite{Zh}:
\begin{equation}\label{eq:MC}
 \mathcal E=\Big\{\frac{X^2}{A}+\frac{Y^2}{B}+\frac{Z^2}{C}=\frac{1}{m}\, \big\vert \, (X,Y,Z)\in \R^3\Big\},
\end{equation}
where $m$ is the mass of the transformed body. Denote by $N=(X_N,Y_N,Z_N)$ the point where the ray defined by the angular momentum of the transformed body $\mathbf{M}=I\Omega$ intersects the ellipsoid \eqref{eq:MC}. Then
\begin{equation*}\label{eq:MC1}
\frac{X_N}{A\Omega_1}=\frac{Y_N}{B\Omega_2}=\frac{Z_N}{C\Omega_3}=\frac{r}{q},
\end{equation*}
where $r=\vert ON\vert$ and  $q^2=\langle I\Omega,I\Omega\rangle$. Using the first of the first integrals of motion \eqref{eq:Zhint}, one gets $q=r\sqrt{2mh}$.

 \begin{figure}[h]\label{fig2}
{\centering{\includegraphics[width=10cm]{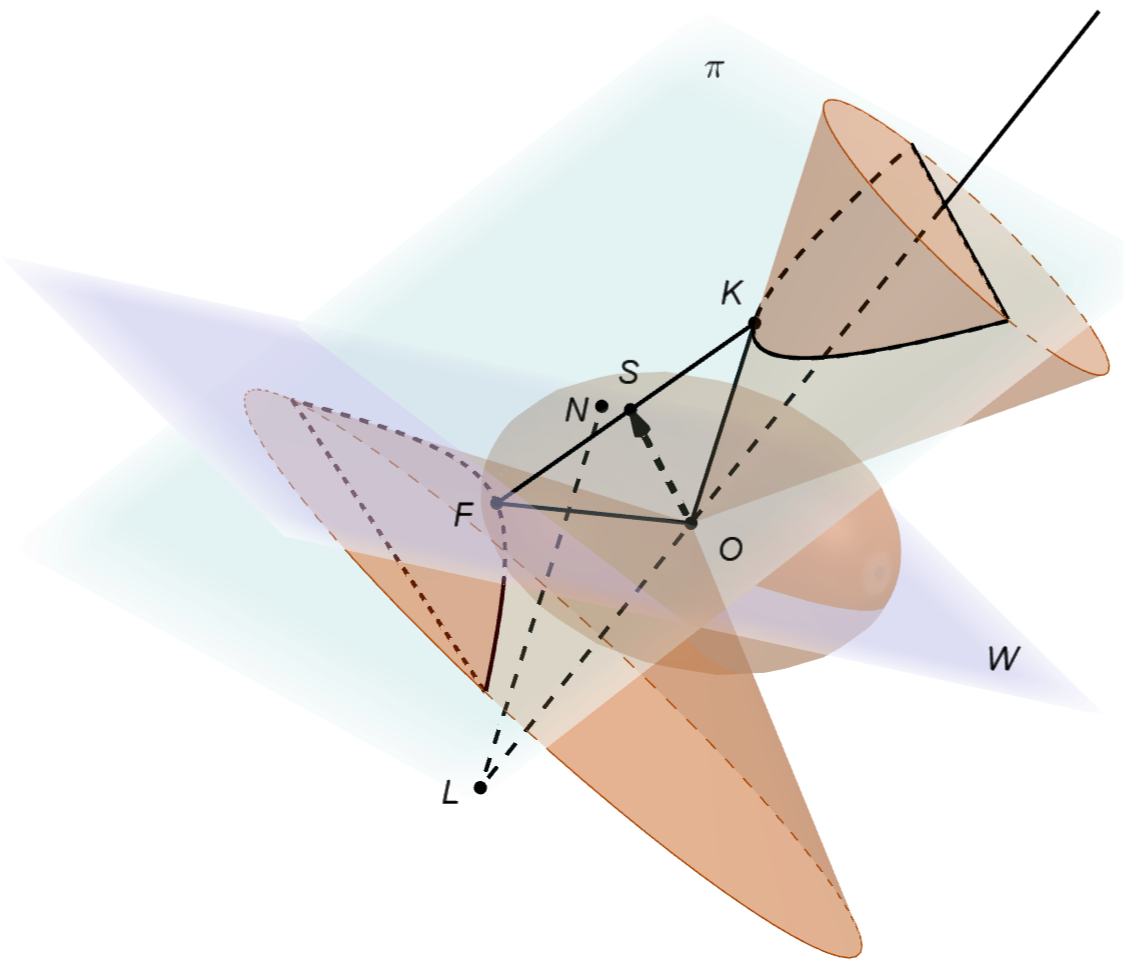}}
\caption{The tangent plane $\pi=T_{N}\mathcal{E}$ of the ellipsoid $\mathcal E$ at the point $N$.}}
\end{figure}

Let $S$ be the foot of the orthogonal line from $O$ onto the tangent plane $\pi=T_{N}\mathcal{E}$ of the ellipsoid $\mathcal E$ at the point $N$  (see Figures 1 and 2), and let  $p=|OS|$. Since the vector $\overrightarrow{OS}$ is proportional to the normal of the ellipsoid at the point $N$:
\[
\Big(\frac{X_N}{A},\frac{Y_N}{B},\frac{Z_N}{C}\Big)=\frac{r}{q}(\Omega_1,\Omega_2,\Omega_3),
\]
it has the direction of the angular velocity of the transformed body. Hence,
\begin{equation}\label{eq:pomega}
p\vert\Omega\vert=\sqrt{\frac{2h}{m}}.
\end{equation}
Let us denote by $L$ the point of the axis of the gyroscope $\ell$, such that $\vert OL\vert={C_2\Omega'_3}/{\sqrt{2mh}}$ and that $\overrightarrow{LO}$ has the same direction as $\mathbf{L}$. One has
\begin{equation}\label{eq:LN}
\overrightarrow{LN}=\overrightarrow{LO}+\overrightarrow{ON}=\frac{1}{\sqrt{2mh}}(\mathbf{L}+I\Omega)=\frac{1}{\sqrt{2mh}}\mathbf{K},
\end{equation}
consequently, the point $N$ lies on the sphere of the radius ${k}/{\sqrt{2mh}}$ with $L$ as the center. The intersection of this sphere with the ellipsoid is the curve $NN'$. We construct three cones: one is $\mathcal{C}_1$ over the curve $NN'$ with the vertex $O$, the second one is $\mathcal{C}_2$ over the curve $NN'$ with the vertex $L$, and the third $\mathcal{C}_3$ is the cone with the vertex $O$ and the directrices being parallel to the directrices of $\mathcal{C}_2$ (see Figures 1 and 2).

Let $W$ be the plane that contains the point $O$ and is orthogonal to the total angular momentum $\mathbf K$.  Let
$\mathbf{k}$, $\hat W$, and $\hat{\mathcal C}_3$ be the total angular momentum $\mathbf K$, the plane $W$, and the cone $\mathcal C_3$ considered in the space frame, respectively. Since $\mathbf k$ is preserved along the motion and $\mathbf K$ belongs to a directrix of the cone $\mathcal C_3$, in the space frame  the body moves in such a way that in any  moment of time, one of the directrices of the cone $\hat{\mathcal{C}}_3$ is perpendicular to the fixed plane $\hat W$ (see Fig. 4).

Thus, if $\mathcal{C}$ is the cone with the vertex $O$ that is orthogonal to $\mathcal{C}_3$ and $\hat{\mathcal C}$ is the corresponding cone
in the space frame, one gets that along the motion, the cone $\hat{\mathcal C}$ rolls with sliding over the plane $\hat W$.
Vice versa, in the body frame, the plane $W$ rolls with sliding over the cone $\mathcal C$.

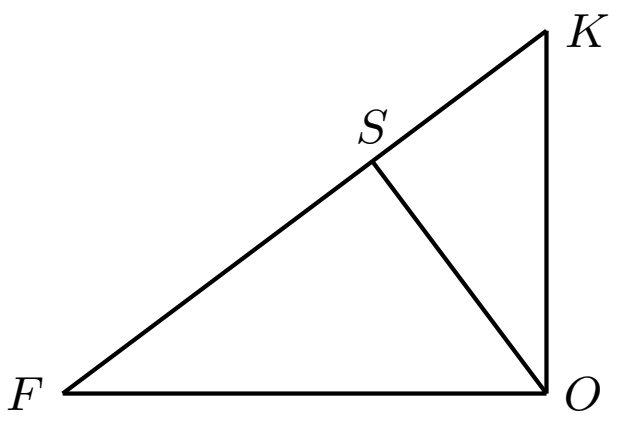
\begin{figure}[h]
\footnotesize{\begin{center}
\begin{tikzpicture}[scale=0.8]
\draw[thick] (0,0) -- (0,3) node [anchor=west] {$K$};
\draw[thick] (0,0)--(-4,0) node [anchor=east] {$F$};
\draw[thick] (0,3)--(-4,0) ;
\draw[thick](0,0) -- (-1.44,1.92) node [anchor=south] {$S$};
\node[anchor=west] (0,0) {$O$};
\end{tikzpicture}
\caption{Triangle $FOK$ in the Zhukovskiy geometric interpretation.}
\end{center}}
\label{fig3}
\end{figure}

Zhukovskiy called $\mathcal{C}_3$ \emph{the cone of the fixed direction}.
To obtain the angular velocities  $\theta'=\theta'(t)$ and $\theta=\theta(t)$ of  rotating and  sliding,  Zhukovskiy decomposed the angular velocity  $\Omega=\Omega(t)=(\Omega_1(t),\Omega_2(t),\Omega_3(t))$ in two
directions  $\overrightarrow{OF}(t)$ and $\overrightarrow{OK}(t)$  respectively.
 The point $K=K(t)\in\mathcal C_3$ is the intersection of the tangent plane  $\pi=T_N\mathcal E$ of the ellipsoid \eqref{eq:MC} at the point $N=N(t)$ with the line that contains $O$ and has the direction $\mathbf K=\mathbf{K}(t)$ (see Fig. 2).
 We denote the intersection of the moving line $KS(t)$ with the plane $W=W(t)$ by $F=F(t)$, and observe that $F$  also belongs to the cone $\mathcal C$. Indeed, the directrix of the cone $\mathcal C$ that corresponds to the directrix $OK$ of the cone $\mathcal C_3$ has the direction equal to the normal of the tangent plane $T_K\mathcal C_3$, which, according to \eqref{MEP1}, is given by
 \[
\mathbf D(t)=\mathbf K\times (\mathbf K\times\Omega)=
(2h+\langle \mathbf L,\mathbf \Omega\rangle) \mathbf K -k^2\Omega\in\Span\{\mathbf K,\mathbf \Omega\}.
 \]
 From the above equation and the definition of the point $F$, we get that $\overrightarrow{OF}$ is proportional to $\mathbf D$.

Therefore, $\triangle FOK$ is a right triangle and the point $S$ belongs to the hypotenuses $KF$ (see Figures 2 and 3).

Let $\alpha$ be the angle $\angle SOK$. Using \eqref{eq:pomega}, one calculates
$\theta=\vert\Omega\vert\cos{\alpha}=\frac{1}{p}\sqrt{\frac{2h}{m}}\cos\alpha$. Similarly, one has $\theta'=\frac{1}{p}\sqrt{\frac{2h}{m}}\sin\alpha$.
Therefore, we obtain:
\[
\theta=\frac{1}{\vert OK\vert}\sqrt{\frac{2h}{m}},\qquad \theta'=\frac{1}{\vert OF\vert}\sqrt{\frac{2h}{m}}.
\]

Finally, we are ready to formulate the Zhukovskiy geometric interpretation: \emph{Let us consider the family of the tangent planes of the ellipsoid \eqref{eq:Zhint}, in all the points of the curve $NN'$ and denote by $c_1$ and $c_2$ the curves of intersection of this family of planes with the cones $\mathcal{C}_3$ and $\mathcal{C}$.  In the space frame, the body moves in such a way that the cone $\hat{\mathcal C}$ rotates with sliding over the fixed plane $\hat W$, so that the angular velocities of  sliding and rotating are inversely proportional to the segments of the directrices of the cones of fixed directions $\mathcal{C}_3$ and the orthogonal cone $\mathcal{C}$ between $O$ and the curves $c_1$ and $c_2$ (see Fig. 4).}

\begin{figure}[h]\label{fig4}
{\centering{\includegraphics[width=8cm]{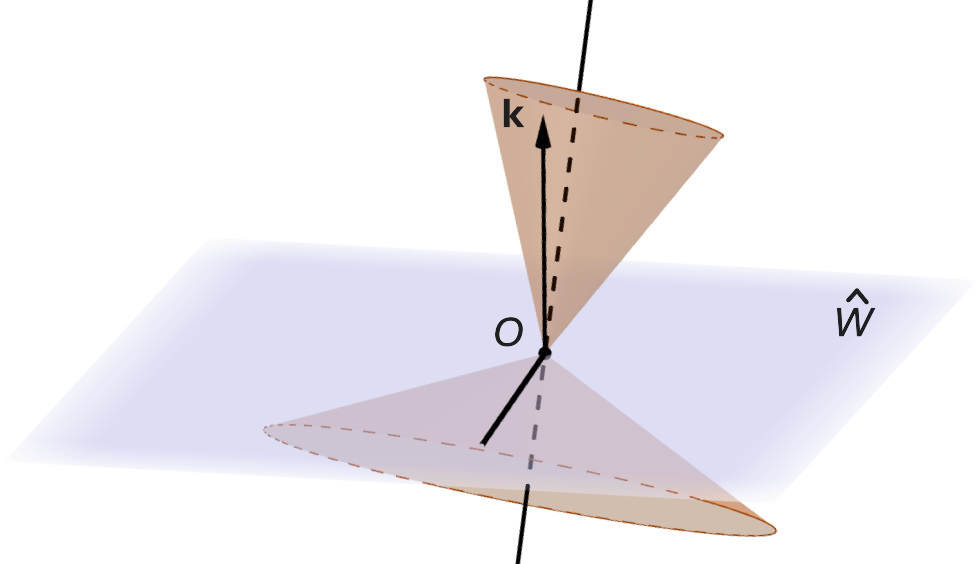}}
\caption{Rolling of the cone $\hat{\mathcal C}$ over the plane $\hat W$ in the  space reference frame.}}
\end{figure}

 It is interesting that Zhukovskiy's initial goal was to consider  motion of a rigid body with cavities filed with a fluid. Then he interpreted these  equations of motion as the equations of motion of the Euler case of a rigid body with a gyroscope \cite{Zh}. For recent contributions to a motion or a rigid body with cavities filed with a fluid see \cite{Ol2025} and reference therein.
Considering the applications, see \cite{DA2025}, where a motion of a satellite within the Earth magnetic field is studied.

\section{Reduction of the Zhukovskiy integrable case to the sphere $S^{n-1}$}\label{sec3}

A straightforward generalization of the system \eqref{MEP} to $so(n)$ is given by
\begin{equation}\label{rMEP}
\dot M=[M+L,\Omega], \qquad \Omega=I^{-1}M,
\end{equation}
representing  motion of a free rigid body  about a fixed point in $\R^n$ with a gyroscope of the angular momentum $L$ in the body frame and the inertia
operator $I\colon so(n)\to so(n)$ of the system body+gyroscope.  It has a form
\[
I(\Omega)=J\Omega+\Omega J,
\]
where $J$ is a symmetric matric - the mass tensor \cite{FK1995}.
An $n$-dimensional rigid body dynamics was extensively studied and we mention just a part of the literature \cite{Ar, Fe1993, FK1995, Ma, BM1999},  while
an addition of a gyroscope to the rigid body system was studied in, e.g., \cite{RST, DGJ2026}.

The Euler equations \eqref{rMEP} are Hamiltonian
with respect to the magnetic Lie-Poisson structure on $so(n)$:
\[
\{M_{ij}, M_{jk}\}_L=-M_{ik}-L_{ik},
\]
with a Hamiltonian function
\begin{equation*}
\label{hf-ha}
H=\frac12\langle M,\Omega\rangle=-\frac14\tr(M\Omega).
\end{equation*}

Here we identify the dual spaces by the invariant scalar product on $so(n)$:
$\langle\xi_1,\xi_2\rangle=-\frac12\tr(\xi_1\xi_2)$.
All skew-symmetric matrices are considered with respect to the standard basis
$\mathbf E_i\wedge \mathbf E_j=\mathbf E_i\otimes \mathbf E_j-\mathbf E_j\otimes \mathbf E_i=\mathbf E_i\mathbf E_j^T-\mathbf E_j\mathbf E_i^T$\footnote{In the matrix notation we consider vectors as column matrices.},
where $[\mathbf E_1,\dots,\mathbf E_n]$ is an orthonormal basis in the body frame
\[
\mathbf E_1=(1,0,\dots,0),\dots, \mathbf E_n=(0,\dots,0,1).
\]

Equivalently, by considering the total angular momentum $K=M+L$ of the system body+gyroscope,  the system \eqref{rMEP} takes the form:
\begin{equation}\label{rEP}
\dot K=[K, \Omega],\qquad \Omega=I^{-1}\big(K-L\big).
\end{equation}

The equations \eqref{rEP} are also Hamiltonian with respect to the standard Poisson brackets,
$\{K_{ij}, K_{jk}\}_0=-K_{ik}$,
and the Hamiltonian function  having a linear in momenta term
\begin{equation}\label{eq:H1}
H_1=\frac12\langle K,I^{-1} K\rangle-\langle K, I^{-1} L\rangle.
\end{equation}

%The mapping $M\mapsto K=M+L$ is a Poisson mapping between the magnetic and the standard Lie-Poisson brackets. Therefore,
%the Casimir functions of the standard structure
%\[
%P_k=\tr(K^{2k}), \qquad k=1,\dots,\Big[\frac{n}2\Big],
%\]
%by setting $K=M+L$, induce the Casimir functions of the magnetic structure.

Following \cite{DGJ2023}, we assume a
symmetric system rigid body+gyroscope with the \emph{Zhukovskiy condition}
\[
I=\sigma\cdot \mathbf{Id}_{so(n)},
\]
implying the simplified equations \eqref{rEP} in the form
\begin{equation}\label{ZhCase}
\dot K=\sigma^{-1}[L,K].
\end{equation}

As a particular case of Theorem 6.1 in \cite{DGJ2026} we obtain that the equations of motion \eqref{ZhCase} are Liouville integrable for a generic $L\in so(n)$, with a complete set of commuting first integrals
$F_{k,l}(K)$ obtained by the argument translation from the invariants (see Mishchenko and  Fomenko \cite{MF1}):
\[
\tr((K+\lambda L)^{2k})=\sum_{l} \lambda^l F_{k,l}(K), \quad k=1,\dots,\Big[\frac{n}2\Big], \quad l=0,\dots,2k.
\]

In the rest of the section we set $L=\frac{s}2\kappa$, $s\in\R$, and $\sigma=1$.
More precisely, we consider the magnetic flow  (see \cite{novikov}) on $T^*SO(n)$ with respect to the canonical symplectic structure and the left-invariant Hamiltonian
with a linear term in the angular momentum:
\[
H_1=\frac12 \langle K,K\rangle -\frac{s}2\langle K, \kappa \rangle.
\]

The Hamiltonian equations on the total phase space $T^*SO(n)$, in the left-trivialization $SO(n)\times so(n)(R,K)$, are given by
\begin{align}
\label{levi1} & \dot K=\frac{s}2[\kappa,K], \\
\label{levi2} & \dot R=R\cdot\Omega, \qquad \Omega=K-\frac{s}2\kappa.
\end{align}

The matrix $R\in SO(n)$ represents the mapping of the body reference frame to the space reference frame. We have
\[
R_{ij}=\langle \mathbf e_i,\mathbf E_j\rangle,
\]
where $[\mathbf e_1,\dots,\mathbf e_n]$ is an orthonormal basis in the space frame.
In other words, the $i$-th row of the matrix $R$ is the vector $\mathbf e_i$ written in the body frame, while the $j$-th column of $R$ is the vector $\mathbf E_j$ written the space frame.

Since $H_1$ is left-invariant, the components of the right-invariant momentum
\begin{equation}\label{Psi}
\texttt{k}=\psi(R,K)=R K R^{-1}
\end{equation}
are first integrals of motion and $\texttt{k}$ represents the total momentum in the space frame.
Note that $\texttt{k}$
is the momentum mapping of the left-action of $SO(n)$ on $T^*SO(n)(R,K)$ endowed with the canonical symplectic structure,
Vice verse, %the linear left-invariant functions on $T^*SO(n)$ can be seen as the momentum mapping of the right $SO(n)$-action.
in the right-trivialization $SO(n)\times so(n)(R,\texttt{k})$ of $T^*SO(n)$, the right action
\begin{equation}\label{desno}
(R,\texttt{k})\longmapsto (RB^{-1},\texttt{k}), \qquad B\in SO(n),
\end{equation}
is Hamiltonian with the momentum mapping
\[
K=\phi(R,\texttt{k})=R^{-1} \texttt{k} R.
\]

Thus, in the right-trivialization we can rewrite the Hamiltonian function $H_1$ as
\begin{equation}\label{collective}
h_1(R,\texttt{k})=\frac12 \langle \phi(R,\texttt{k}),\phi(R,\texttt{k})\rangle -\frac{s}2\langle \phi(R,\texttt{k}), \kappa \rangle.
\end{equation}

The Euler equation \eqref{levi1} is the Poisson reduction of the system \eqref{levi1}, \eqref{levi2}
with respect to the left $SO(n)$-action:  $(T^*SO(n))/SO(n)\cong so(n)$.
On the other hand, we can perform the symplectic Marsden-Weinstein reduction of the left action
\[
(R,K)\longmapsto (AR,K), \qquad A\in \hat{SO}(n-1)\subset SO(n),
\]
of the subgroup
\[
\hat{SO}(n-1)=\{A=
 \left(
\begin{array}{cc}
1 & 0  \\
0 & S
\end{array} \right)\, \vert\, S\in SO(n-1)\}\cong SO(n-1).
\]

We write simply $SO(n-1)$ instead of $\hat{SO}(n-1)$.
From \eqref{Psi}, the momentum mapping of the left ${SO}(n-1)$-action is given by
\[
\Psi=\pr_{{so}(n-1)}\psi=
\pr_{{so}(n-1)}(R K R^{-1})=\pr_{{so}(n-1)}(\texttt{k})=\texttt{k}-(\texttt{k} \mathbf E_1)\wedge \mathbf E_1.
\]

We have that $\Psi=0$ if and only if $\texttt{k}=(\texttt{k} \mathbf E_1)\wedge \mathbf E_1$, that is,
\[
\texttt{k}_{ij}=0, \qquad 1\le i,j\le n-1.
\]

The symplectic reduction at the zero-level set (in the right-trivialisation)
\[
(T^*SO(n))_0=\Psi^{-1}(0)=\{(R,\texttt{k})\in T^*SO(n)\, \vert\, \texttt{k}_{ij}=0, \, 1\le i,j\le n-1\}
\]
is symplectomorphic to the cotangent bundle of the sphere
\[
S^{n-1}=SO(n)/SO(n=1), \qquad  (T^*SO(n))_0/SO(n-1)\cong T^*S^{n-1}.
\]

 A general framework for a symplectic reduction of  Hamiltonian systems with symmetries on magnetic cotangent bundles is given in \cite{KNP2005}.
Since the magnetic field that we consider here is exact, we work within the standard symplectic reduction of cotangent bundles, additionally employing the fact that the Hamiltonian function \eqref{collective} is the
pull-back of the Hamiltonian function on $so(n)$ by using the momentum mapping of the right $SO(n)$-action, which is also known as a collective Hamiltonian function, see \cite{GS}.

As usual, we consider the cotangent bundle of the sphere  defined by:
\[
T^*S^{n-1}=\{(\gamma,p)\in\R^{2n}\, \vert\, \langle \gamma,\gamma\rangle=1, \, \langle \gamma,p\rangle=0\}.
\]

The canonical symplectic form  $\varpi$ on $T^*S^{n-1}$ is the restriction of the canonical
symplectic form  $\omega=\sum_{i=1}^n dp_i\wedge d\gamma_i$ from the  symplectic linear space $\R^{2n}(\gamma,p)$:
\[
\varpi=\sum_{i=1}^n dp_i\wedge d\gamma_i\vert_{T^*S^{n-1}}.
\]

The vector $\gamma\in\R^n$ represents the vector $\mathbf e_1$ considered in the body frame:
$\gamma_i=\langle \mathbf e_1,\mathbf E_i\rangle$, $i=1,\dots,n$. Indeed,
we have a natural identification
\[
(\gamma_1,\dots,\gamma_n)=(\langle \mathbf e_1,\mathbf E_1\rangle,\dots,\langle \mathbf e_1,\mathbf E_n\rangle) \longleftrightarrow \{AR\,\vert\,A\in SO(n-1)\}.
\]

At the level of configuration space $SO(n)$ we have the commutativity of the left $SO(n-1)$-action and the right $SO(n)$-action:
\begin{equation*}\label{dijagram}
\begin{tikzcd}
 R \arrow[r, maps to] \arrow[d, maps to]  & AR \arrow[d, maps to] \\
RB^{-1} \arrow[r, maps to] &   ARB^{-1}
\end{tikzcd}, \qquad R\in SO(n), \, A\in SO(n-1), \, B\in SO(n).
\end{equation*}

Therefore, the extensions of the left $SO(n-1)$-action and the right $SO(n)$-action to the cotangent bundle $T^*SO(n)$ commute between themselves as well.

The right $SO(n)$-action on $(T^*SO(n))_0\subset T^*SO(n)$  (see \eqref{desno}) induces
the standard Hamiltonian $SO(n)$-action on $(T^*S^{n-1},\varpi)$
\[
(\gamma,p)\longmapsto (B\gamma,Bp), \qquad B\in SO(n),
\]
with the momentum mapping (modulo the sign convention):
\begin{equation}\label{Phi}
\Phi(\gamma,p)=\gamma\wedge p, \qquad  \text{i.e.,} \qquad  \Phi_{ij}=\gamma_ip_j-\gamma_j p_i, \,\, i,j=1,\dots,n.
\end{equation}

Further, let $\pi\colon (T^*SO(n))_0\to (T^*SO(n))_0/SO(n-1)=T^*S^{n-1}$ be the natural projection. With the above choice of the sign in the momentum mapping $\Phi$, it coincides with the induced map from the
restriction of $\phi$ to the zero-set $(T^*SO(n))_0$:
\begin{equation*}
\Phi(\gamma,p)=\phi(R,\texttt{k}), \qquad \text{where} \qquad (R,\texttt{k})\in \pi^{-1}(\gamma,p)\subset (T^*SO(n))_0.
\end{equation*}

 Therefore, the Hamiltonian $h_1$  (see \eqref{collective}) induces the reduced Hamiltonian function on $T^*S^{n-1}\subset \R^{2n}$ given by:
\begin{equation*}
\begin{aligned}
h_{red}(\gamma,p)=&\frac12\langle \Phi,\Phi\rangle-\frac{s}2\langle \Phi,\kappa\rangle
=\frac12\langle p,p\rangle\langle\gamma,\gamma\rangle-\frac12\langle p,\gamma\rangle^2+\frac14 s\tr\big((\gamma\otimes p-p\otimes \gamma)\kappa\big)\\
=&\frac12\langle p,p\rangle\langle\gamma,\gamma\rangle-\frac12\langle p,\gamma\rangle^2+\frac{s}2\langle p,\kappa\gamma\rangle.
\end{aligned}
\end{equation*}

Note that the canonical Hamiltonian flow of the function $h_{red}$ in  the ambient space $(\R^{2n}(\gamma,p),\omega)$:
\begin{align*}
&\dot \gamma=\frac{\partial h_{red}}{\partial p}=p\langle \gamma,\gamma\rangle -\langle p,\gamma\rangle\gamma +\frac{s}2\kappa\gamma,\\
&\dot p =-\frac{\partial h_{red}}{\partial \gamma}=\frac{s}2\kappa p -\langle p,p\rangle \gamma+\langle \gamma,p\rangle p.
\end{align*}
has the constraint functions $\langle \gamma,\gamma\rangle$ and $\langle \gamma,p\rangle$ as first integrals. Thus, we get:

\begin{thm}\label{prva}
The Marsden-Weinstein $SO(n-1)$--reduced system of the $n$-dimensi\-onal Zhukovskiy integrable case \eqref{levi1}, \eqref{levi2} at the zero-value of the momentum mapping is given by the magnetic flow on $T^*S^{n-1}$
\begin{equation}\label{MM}
\dot \gamma=p+\frac{s}2\kappa\gamma, \qquad \dot p =\frac{s}2\kappa p -\langle p,p\rangle \gamma
\end{equation}
with respect to the canonical symplectic form $\varpi$ and the Hamiltonian function with a linear term in momenta
$h_{red}=\frac12\langle p,p\rangle+\frac{s}2\langle p,\kappa\gamma\rangle$.
\end{thm}

\begin{rem}
 In the Lagrangian formulation, the equations of motion \eqref{MM} are equivalent to the equations \eqref{novi}. They represent motion of a
material point of a unit mass on the sphere under the influence of the Neumann potential
\[
V=\frac18\langle \gamma, s^2\kappa^2\gamma\rangle
\]
and the constant magnetic field $s\kappa$.
\end{rem}

Note that the time derivative of the momentum mapping $\Phi$ along the flow \eqref{MM} is:
\[
\dot\Phi=\frac{s}2[\kappa,\Phi],
\]
which is in agreement with the Euler equation \eqref{levi1} and leads to the following statement:

%By introducing the variable $p=u-s\kappa\gamma$, we can write the equations
%in the form
%\begin{align}
%\label{MM1} &\dot \gamma=p,\\
%\label{MM2} &\dot p =-s\kappa(p+s\kappa\gamma)+\lambda_1\gamma=-s\kappa p-s^2\kappa^2\gamma+s\lambda_1 \kappa^2\gamma+\lambda_1\gamma-s\kappa p
%\end{align}

\begin{cor}\label{posledica} Let $so(n)_\kappa=\{\xi\in so(n)|[\xi,\kappa]=0\}$ be the isotropy subalgebra of $\kappa$  within $so(n)$.
Then the components of
\begin{equation}\label{neterini}
 \mathcal J=\pr_{so(n)_\kappa}\Phi=\pr_{so(n)_\kappa}\gamma \wedge p
\end{equation}
are  first integrals of the equations of motion \eqref{MM}, where the projection is considered with respect to the invariant scalar product $\langle\cdot,\cdot\rangle$
on $so(n)$.
\end{cor}

Let $\ell:=[n/2]$ and consider a basis $[\mathbf E_1,\dots,\mathbf E_n]$ of $\R^n$ in which the matrix $\kappa\in so(n)$ is given by:
\begin{equation}\label{kappa}
\kappa=\kappa_{12} \mathbf E_1\wedge \mathbf E_2+\kappa_{34} \mathbf E_3\wedge \mathbf E_4 +\dots +\kappa_{2\ell-1,2\ell}\mathbf E_{2\ell-1}\wedge \mathbf E_{2\ell},
\end{equation}
where $\kappa_{2i-1,2i}\ge 0, \, i=1,\dots,\ell$.
Then, from Corollary \ref{posledica}, we get the Noether first integrals:
\begin{equation}\label{linearni}
\mathcal J_i=\Phi_{2i-1,2i}=\gamma_{2i-1}p_{2i}-\gamma_{2i}p_{2i-1},
\qquad i=1,\dots,\ell.
\end{equation}

Recall that $\kappa$ is a \emph{regular} element of $so(n)$ if
$so(n)_\kappa=\Span\{\kappa,\kappa^3,\dots,\kappa^{2\ell-1}\}$  i.e., the adjoint isotropy group $SO(n)_\kappa$ is a maximal torus.
In particular,  $\kappa$ of the form \eqref{kappa} is regular if $\kappa_{2i-1,2i}\ne\kappa_{2j-1,2j}$, and for $n$ odd,
in addition, all $\kappa_{2i-1,2i}$ should be different from zero.
If \eqref{kappa} is regular, then the set of functions \eqref{linearni} gives all Noether first integrals.

Furthermore, for $n=2\ell$, following \cite{DGJ2025b},
 it is convenient to use a complex notation.
We set
\[
z_i=\gamma_{2i-1}+\sqrt{-1}\gamma_{2i}, \qquad w_i=p_{2i-1}+\sqrt{-1} p_{2i}, \qquad i=1,\dots,\ell.
\]
Then the equations of motion \eqref{MM} take the form:
\begin{equation}
\label{comMM}
\dot z=w-\frac{s}2\sqrt{-1}Kz, \qquad
\dot w=-\frac{s}2\sqrt{-1}K w+\mu z, \qquad \mu=-\langle w,\bar w\rangle,
\end{equation}
where $z=(z_1,...,z_{\ell})$, $w=(w_1,...,w_{\ell})$ and
\begin{equation}\label{KK}
K=\diag(\kappa_{1,2},...,\kappa_{2\ell-1,2\ell}).
\end{equation}

Since the $n$-dimensional Zhukovskiy case \eqref{levi1}, \eqref{levi2} is completely integrable, for a generic $\xi\in\psi(T^*SO(n))\subset so(n-1)$, when $SO(n-1)_\xi$ is a maximal torus, the reduced system on $\psi^{-1}(\xi)/SO(n-1)_\xi$ is  also completely integrable \cite{Jo2008, Zu}. In our case, the isotropy group $SO(n-1)_0$ is the whole $SO(n-1)$ and  we present a direct proof of the complete integrability of the reduced system.

\begin{thm}\label{druga}
The equations of motion \eqref{MM}  imply a Lax representation:
\begin{equation}
\dot L(\lambda)=[L(\lambda), A],\qquad L(\lambda)=\Phi+\lambda \kappa,\qquad A=-\frac{s}2\kappa,
\label{matrix}
\end{equation}
while the equations \eqref{comMM} imply a Lax representation:
\begin{equation}
\dot{\mathcal L}(\lambda)=[\mathcal L(\lambda),\mathcal A],\qquad \mathcal L(\lambda)=\Phi^{u(\ell)}+\lambda \sqrt{-1}K,\qquad \mathcal A=\frac{s}2\sqrt{-1} K.
\label{matrix2}
\end{equation}
Here $\Phi$ is the momentum mapping \eqref{Phi}, $K$ is given by \eqref{KK}, and $\Phi^{u(\ell)}$ is the momentum mapping of the standard  $U(\ell)$ action on
$T^*S^{2\ell-1}$:
\[
\Phi^{u(\ell)}=\frac{1}{2}(w\otimes \bar z - z\otimes \bar w).
\]
\end{thm}

From the Lax representation \eqref{matrix} and \eqref{matrix2}, we have the involuting first integrals:
\[
\tr\big(L(\lambda)^{2k}\big) \quad \text{and} \quad \tr\big(\mathcal L(\lambda)^{k}\big),  \qquad k=1,\dots,\ell.
\]

For a regular $\kappa$ of the form \eqref{kappa} and $n=2\ell$, we get
the first integrals quadratic in momenta:
\begin{equation}\label{integrali}
\mathcal I_k=\sum_{1\le i<j\le\ell}\frac{\kappa_{2i-1,2i}^k-\kappa_{2j-1,2j}^k}{{\kappa_{2i-1,2i}-\kappa_{2j-1,2j}}}\big((\Phi_{2i-1,2j}-\Phi_{2i,2j-1})^2+(\Phi_{2i-1,2j-1}+\Phi_{2i,2j})^2 \big),
\end{equation}
$k=1,\dots,\ell-1$. The polynomials $\mathcal I_1,\dots,\mathcal I_\ell$ as expressions for the first integrals
follow from the Lax representation
\eqref{matrix2}, which allows for a more compact form of the first
 integrals than the Lax representation \eqref{matrix}.

The integrals \eqref{integrali} are independent, and together with the Noether first integrals \eqref{linearni} form a
complete set of commuting first integrals on $T^*S^{2\ell-1}$. Let us note that  the first integrals \eqref{linearni}, \eqref{integrali} are the same as the ones given by Bolsinov in the construction of integrable geodesic flows on the sphere $S^{2\ell-1}\subset \R^{2\ell}$ represented as the homogeneous space $U(\ell)/U(\ell-1)$ (see Appendix C, Section 3.3, \cite{BM2003}).

In the case $n=2\ell-1$, we set $\kappa_{2\ell-1,2\ell}=0$ and note that the manifold
$\{\gamma_{2l}=0,p_{2l}=0\}$ is invariant under the flow of \eqref{MM} within $T^*S^{2\ell-1}$, as in \cite{DGJ2025b}. We obtain $2\ell-2$ commuting first integrals on $T^*S^{2\ell-2}$: the Noether integrals $\mathcal J_1,\dots,\mathcal J_{\ell-1}$ and the restrictions of $\mathcal I_1,\dots,\mathcal I_{\ell-1}$ to $\{\gamma_{2l}=0,p_{2l}=0\}$.

Therefore, we obtain:

\begin{thm}\label{treca}
If $\kappa$ is regular element of $so(n)$ then the magnetic flow \eqref{MM} is Liouville integrable.
\end{thm}

 Otherwise, when $\kappa$ is singular, it can be proved that  the system \eqref{MM} is integrable in the non-commutative sense by means of the additional Noether first integrals, which are the components of \eqref{neterini}, and a suitable linear combinations of integrals \eqref{integrali}.

\begin{rem}
Let us note that the reduced system \eqref{MM} is slightly different from the magnetic geodesic flow obtained as the reduced system of Demchenko's nonholonomic problem of rubber rolling of a ball with a gyroscope \cite{DGJ2023, DGJ2025}:
\begin{equation}\label{magGF}
\dot\gamma=  p,\qquad \dot p  ={s}\kappa p+\big({s}\langle p,\kappa\gamma \rangle-\langle p,p\rangle\big)\gamma.
\end{equation}

The system \eqref{magGF} is Hamiltonian with respect to the twisted symplectic form
\[
\varpi+s\sum_{i<j}\kappa_{ij} d\gamma_i\wedge d\gamma_j\vert_{T^*S^{n-1}}
\]
and with the Hamiltonian function $h=\frac12\langle p,p\rangle$. The magnetic geodesic flow \eqref{magGF}
is completely integrable with a Neumann-type Lax representation (see \cite{DGJ2025, DGJ2025b, BKM2025}), but with the same set of the Noether first integrals
as the magnetic system \eqref{MM} (see \cite{DGJ2025b}).
\end{rem}

\section{Reduction of the Zhukovskiy--Volterra gyroscope to the sphere $S^2$}\label{sec4}

In the $3$-dimensional case, the Zhukovskiy-Volterra gyroscope \eqref{MEP} (or \eqref{rMEP}) is integrable for an arbitrary inertia operator $I$.
By using the reduction procedure described in Section \ref{sec3} for the Hamiltonian function with a linear in momenta term
\[
H_1=\frac12\langle \mathbf K,I^{-1} \mathbf K\rangle-\langle \mathbf K, I^{-1} \mathbf L\rangle,
\]
we get the reduced Zhukovskiy-Volterra system on the cotangent bundle $T^*S^2$ with the Hamiltonian function
\begin{equation}
\label{ZVred}
\begin{aligned}
h_{ZV,red}(\gamma,p)=&\frac12\langle  p \times \gamma, I^{-1}(p\times \gamma)\rangle-\frac{s}2\langle \vec{\kappa}, p \times \gamma \rangle,
\end{aligned}
\end{equation}
where $s\vec{\kappa}:=2I^{-1}(\mathbf L)$ and $s$ is a real parameter. As above, $\gamma$ represents the position of the vector $\mathbf e_1$ fixed in the space and considered in the
body frame. The subgroup $SO(2)\subset SO(3)$ is the group of rotations about $\mathbf E_1$, and the reduction is taken for the zero value of the corresponding momentum
mapping.

As in Section \ref{sec3}, we consider the canonical Hamiltonian equations of $h_{ZV,red}$ in  $(\R^{6}(\gamma,p),\omega)$:
\begin{align}
\label{MM2a} &\dot \gamma=\frac{\partial h_{ZV}}{\partial p}=\gamma\times I^{-1}(p\times \gamma)+\frac{s}2\vec{\kappa}\times\gamma,\\
\label{MM2b} &\dot p =-\frac{\partial h_{ZV}}{\partial \gamma}=p\times I^{-1}(p\times \gamma)+\frac{s}2\vec{\kappa}\times p.
\end{align}

Again, the flow has the constraint functions $\langle \gamma,\gamma\rangle$ and $\langle \gamma,p\rangle$ as its first integrals.
 The time derivative of the momentum mapping $\vec\Phi=p\times\gamma$ along the flow \eqref{MM2a}, \eqref{MM2b} is
\[
\dot{\vec\Phi}=\vec\Phi\times \big(I^{-1}\vec\Phi-\frac{s}{2}\vec\kappa\big),
\]
which is in agreement with the  equation \eqref{MEP1} and leads to the first integral
\[
\mathcal I(\gamma,p)=\langle \vec\Phi,\vec\Phi\rangle=\frac12\langle p,p\rangle\langle\gamma,\gamma\rangle-\frac12\langle p,\gamma\rangle^2=\langle p,p\rangle \vert_{T^*S^2}.
\]

Thus, we obtain:

\begin{thm}\label{cetvrta}
The Marsden-Weinstein $SO(2)$-reduced system of the Zhukovskiy-Volterra gyroscope is the magnetic flow on $T^*S^{2}$, given by equations \eqref{MM2a}, \eqref{MM2b}.
The system is Hamiltonian  with respect to the canonical symplectic form $\varpi$ and the Hamiltonian function \eqref{ZVred}.
It is completely integrable with a first integral quadratic in momenta
$
 \mathcal I(\gamma,p)=\langle p,p\rangle.
$
\end{thm}

It is also convenient to have a Lagrangian description of the dynamics.
To that end, we need to perform the inverse Legendre transformation of the Hamiltonian function for the mechanical systems with constraints. By definition (see \cite{Ar}), the
Lagrangian is defined by
\[
L_{ZV}(\gamma,\dot\gamma)=\langle p, \dot\gamma\rangle -H(\gamma,p)\vert_{p=p(\gamma,\dot\gamma)},
\]
where $p=p(\gamma,\dot\gamma)$ is the unique solution of the equation \eqref{MM2a} that satisfies the constraint $\langle p,\gamma\rangle=0$.
By using the identity
\[
I^{-1}(p\times\gamma)=\frac{1}{\det I} Ip\times I\gamma,
\]
we obtain
\begin{equation}\label{pGamma}
p(\gamma,\dot\gamma)=\frac{\det I}{\langle I\gamma,\gamma\rangle}I^{-1}\dot\gamma+\frac{s}2\gamma\times I\vec{\kappa}+\frac{s}{2}\frac{\langle I\vec{\kappa},\gamma\rangle}{\langle I\gamma,\gamma\rangle} I\gamma\times\gamma-\frac{\det I}{\langle I\gamma,\gamma\rangle}\langle I^{-1}\dot{\gamma},\gamma\rangle\gamma.
\end{equation}

As a result, by subtracting the constant $\frac{s}8\langle I\vec\kappa,\vec\kappa\rangle$, we get:

\begin{thm}\label{peta}
The Lagrangian of the reduced  Zhukovskiy-Volterra gyroscope on $S^{2}$ is given by
\[
L_{ZV}(\gamma,\dot\gamma)=\frac{\det I}2\frac{\langle I^{-1}\dot\gamma,\dot\gamma\rangle}{\langle I\gamma,\gamma\rangle}+\frac{s}{2}\big\langle \dot\gamma, \gamma\times I\vec\kappa+\frac{\langle I\kappa,\gamma\rangle}{\langle I\gamma,\gamma\rangle}I\gamma\times\gamma\big\rangle-\frac{s^2}{8}\frac{\langle I\vec\kappa,\gamma\rangle^2}{\langle I\gamma,\gamma\rangle}.
\]
\end{thm}

From \eqref{pGamma} we also retrieve the tangent bundle form of the
first integral: $\mathcal I(\gamma,\dot\gamma)=\langle p(\gamma,\dot\gamma),p(\gamma,\dot\gamma)\rangle$.
Further, for $s=0$, we reobtain the well known expression for the Poisson sphere metric (see \cite{BKF}, where the matrix $I$ corresponds to our inverse inertia operator $I^{-1}$):
\[
ds^2=\det I\frac{\langle I^{-1} d\gamma,d\gamma\rangle}{\langle I\gamma,\gamma\rangle}=ABC \frac{A^{-1}d\gamma_1^2+B^{-1}d\gamma_2^2+C^{-1}d\gamma_3^2}{A\gamma_1^2+B\gamma_2^2+C\gamma^2_3}.
\]

Note that in \cite{Jo2025}, integrable magnetic geodesic flows with respect to the contact form on rank two Stiefel manifolds $V_{n,2}$ were considered,
where, as a special case for $n=3$, under the diffeomorphism $SO(3)\cong V_{3,2}$, it was obtained the Zhukovskiy--Volterra gyroscope. Also, the geometry of the Zhukovskiy--Volterra gyroscope on the covering $S^3$ of the configuration space $SO(3)$, for $A=B$, was studied in details in \cite{IM2024}.
 A different, algebro-geometric construction, which relates the
Zhukovskiy--Volterra gyroscope to the integrable XYZ spin chain is given in \cite{LOZ2006} and \cite{MZ2026}.

The magnetic systems on Lie groups with respect to non-exact magnetic $2$-forms were considered in \cite{MS, T}.

\subsection*{Acknowledgements}
 This research has been supported by the grant IntegraRS of the Science Fund
of Serbia, the Ministry for Education, Science, and Technological Development of Serbia, and the Simons
Foundation grant no. 854861.

\end{document}